\documentclass[twocolumn,aps,superscriptaddress]{revtex4-1}
\usepackage{palatino}
\usepackage[latin1]{inputenc}
\usepackage{epsf}
\usepackage{amsmath,amssymb}
\usepackage{latexsym}
\usepackage{calc}
\usepackage{color}
\usepackage{graphicx}

\newcommand{\ben}{\begin{equation}}
\newcommand{\een}{\end{equation}}
\newcommand{\bea}{\begin{eqnarray}}
\newcommand{\eea}{\end{eqnarray}}

\def\sss{\scriptscriptstyle\rm}

\def\1s{_{1,\sss S}}
\def\2s{_{2,\sss S}}

\def\eps{{\epsilon}}

\begin{document}
\title{Exact Potential Energy Surface for Molecules in Cavities}
\author{Lionel Lacombe}
\affiliation{Department of Physics and Astronomy, Hunter College of the City University of New York, 695 Park Avenue, New York, New York 10065, USA}
\author{Norah M. Hoffmann}
\affiliation{Max Planck Institute for the Structure and Dynamics of Matter and
Center for Free-Electron Laser Science and Department of Physics,
Luruper Chaussee 149, 22761 Hamburg, Germany}
\affiliation{Department of Physics and Astronomy, Hunter College of the City University of New York, 695 Park Avenue, New York, New York 10065, USA}
\author{Neepa T. Maitra}
\affiliation{Department of Physics and Astronomy, Hunter College  of the City University of New York, 695 Park Avenue, New York, New York 10065, USA}
\affiliation{Physics Program and Chemistry Program, Graduate Center of the City University of New York, New York, USA}

\date{\today}
\pacs{}
\begin{abstract}
We find and analyze the exact time-dependent potential energy surface driving the proton motion for a model of cavity-induced suppression of proton-coupled electron-transfer. We show how, in contrast to the polaritonic surfaces, its features directly correlate to the proton dynamics  and  discuss cavity-modifications of its structure responsible for the suppression. The results highlight the interplay between non-adiabatic effects from coupling to photons and coupling to electrons, and suggest caution is needed when applying traditional dynamics methods based on polaritonic surfaces. 

\end{abstract}

\maketitle
Impressive experimental advances~\cite{E16,BGA18,VL18,BWCAS18,HSH18} have led to a rekindling of interest in cavity quantum electrodynamics.  Rapidly expanding applications to molecules and nanostructures require us to go beyond the simplest few-level--single-mode models explored in the early days of quantum mechanics,  with the interplay of coupled electronic, nuclear, and photonic excitations revealing a plethora of new phenomena from enhanced conductivity in semiconductors  to photochemical suppression of chemical reactions  to cavity-enhanced superconductivity to superradiance, see e.g. Refs.~\cite{Orgiu15,FRAR17,KBM16,KM17,HS16,GG19,SRR18}. 
There is now the possibility to manipulate real matter with cavity parameters providing tunable dials for photo-chemical control of reactions, replacing shaped laser pulses as photonic reagents~\cite{E16,RMDCY18,CHVK19}. The hope is attain strong light-matter coupling and control without large power sources, possibly reducing unintended byproducts such as multiphoton absorption and ionization channels. 
 
The cavity clearly modifies the potential that the matter evolves in, and various constructs have been put forward to serve in lieu of the Born-Oppenheimer (BO) surfaces that have proved so instrumental for our understanding of cavity-free dynamics. In particular, ``polaritonic surfaces" that arise from diagonalizing the electron-photon Hamiltonian parametrized by nuclear coordinates have been instructive in interpreting some of the novel phenomena mentioned above~\cite{GGF15,GGF16,FGG18,SRAR19,RTFAR18}.  Another construct are the ``cavity-BO surfaces" where the photonic displacement-field coordinates are treated on the same footing as the nuclear coordinates~\cite{FARR17,FRAR17}.
A complete dynamical picture of how the electronic and photonic degrees of freedom influence the nuclear dynamics can only be obtained when several of such surfaces in the chosen manifold together with their couplings are considered: quite typically at a given time the nuclear wavepacket locally straddles several surfaces, or distinct parts of the nuclear wavepacket are associated with different surfaces. To go beyond using the surfaces only for qualitative interpretation, and to implement them in dynamics schemes, couplings between the surfaces must be included~\cite{KBM16,LFTG17}, and there is interplay between non-adiabatic effects arising from photon-matter coupling and electron-nuclear coupling. Practical necessity calls for approximations which usually work best when this choice of surfaces in some sense represents a ``zeroth order" picture. The situation somewhat mirrors that of the molecule driven by classical light, where, for example in surface-hopping schemes in some situations Floquet states  (which are the classical-light analogues to the polaritonic surfaces) work best~\cite{FLSS11,FHGS17} while in other cases quasi-static (a.k.a. phase-adiabatic, or instantaneous BO) states have been argued to be more appropriate~\cite{TIW96,SKKF03}.

The exact factorization (EF) approach bypasses these questions while at the same time shedding light on them. Originally presented for coupled electron-nuclear systems, a single time-dependent potential energy surface (TDPES) replaces the manifold of static potential energy surfaces, and represents the exact potential that the nuclear wavepacket evolves in, which exactly contains the effects of coupling to the electrons~\cite{AMG10,AMG12}. Generalizations of EF have been made  to include photons~\cite{HARM18,AKT18}; explicit examples of how the coupling to photons affects features of the potential driving an electron are given in Ref.~\cite{AKT18} while Ref.~\cite{HARM18} finds the exact photon-matter coupling-induced corrections to the potential driving the photons. So far, how the presence of the cavity modifies the exact potential driving the nuclei has not been explored.
In this paper, 
we find the exact cavity-modified TDPES for a model that demonstrates suppression of photo-induced proton-coupled electron transfer (PCET), a key process in energy conversion in biological and chemical systems. In contrast to the polaritonic surfaces, its features alone indicate the suppression phenomenon, and it provides the exact, unambiguous force on the nuclei to be used in mixed quantum-classical methods.

\begin{figure}
\includegraphics[width=0.75\columnwidth]{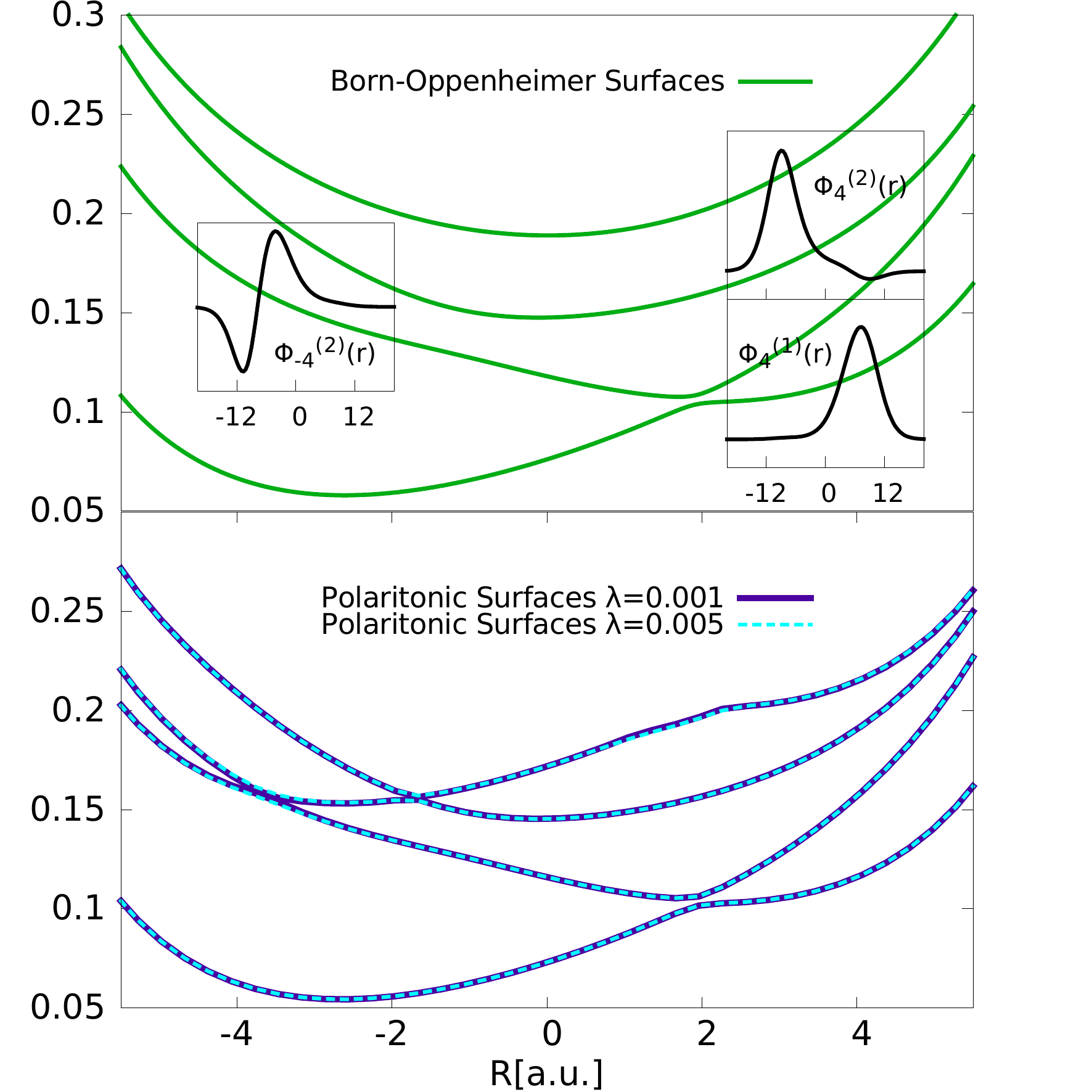}
\caption{Upper panel: The lowest BO surfaces for the PCET model. The initial conditional electronic wavefunction associated with the initial excitation on the donor side is shown in the inset on the left, while those associated with the BO states after a proton transfer are shown on the right. Lower panel: The polaritonic surfaces, for coupling strengths indicated. }
\label{fig:bopolpcet}
\end{figure}

The minimal model of Ref.~\cite{SM95,FH97,FH97b} has proved to be remarkably instructive in studying non-adiabatic effects in cavity-free PCET~\cite{FH97,FH97b,AASG13,AASMMG15}.
The Hamiltonian for the cavity-free matter (one electron and one proton moving between two fixed heavy ions separated by $L$) is
\ben
\hat{H}_m = \hat{T}_n + \hat{H}_{\rm BO} = \hat{T}_n + \hat{T}_e + \hat{V}_m
\een
where $\hat{T}_n = -\frac{1}{2M}\frac{\partial^2}{\partial R^2}$,  $\hat{T}_e = -\frac{1}{2}\frac{\partial^2}{\partial r^2}$, and
\ben
\hat{V}_{m} = \sum_{\sigma = \pm 1}\left(\frac{1}{\vert R +\frac{\sigma L}{2}\vert} - \frac{{\rm erf}\left(\frac{\vert r + \frac{\sigma L}{2}}{a_\sigma}\right)}{\vert r + \frac{\sigma L}{2}\vert}\right) - \frac{{\rm erf}\left(\frac{\vert R - r\vert}{a_f}\right)}{\vert R - r\vert}
\een
 where we have chosen parameters $L = 19.0$a.u.,  $a_+ = 3.1$a.u., $a_- =4.0$a.u.,$a_f = 5.0$a.u.,  and $M = 1836$a.u., the proton mass. Atomic units, in which $\hbar = e^2 = m_e = 1$, are used here and throughout.
The top panel in Fig.~\ref{fig:bopolpcet} shows the BO surfaces for the system outside the cavity. Considering an initial sudden vertical electronic excitation out of the ground-state donor well on the left to the first excited BO state, the nuclear wavepacket slides down the surface and splits soon after encountering the avoided crossing (see the figures shortly and movie in the Supplemental Material). The part of the nuclear wavepacket evolving on the lower surface then becomes associated with an electron transfer as evident from comparing the conditional BO electronic wavefunctions shown in the insets in Fig.~\ref{fig:bopolpcet}. To investigate how placing the molecule in a cavity affects the PCET, we consider the non-relativistic photon-matter Hamiltonian in the dipole approximation in the Coulomb gauge~\cite{T13,FARR17,RFPATR14,RTFAR18,HARM18}
\ben
\hat{H}  = \hat{H}_m + \hat{H}_p + \hat{V}_{pm} + \hat{V}_{\rm dipSE}
\label{eq:fullH}
\een
where, considering a single cavity-mode of frequency $\omega_\alpha$,
\ben
\hat{H}_p(q)= \frac{1}{2}\left(\hat{p}_\alpha^2 + \omega_\alpha^2\hat{q}_\alpha^2\right)\; {\rm and} \; \hat{V}_{pm}= \omega_\alpha\lambda_\alpha\hat{q}_\alpha\left(R - r\right)
\een
where $\hat{q}_\alpha = \sqrt{\frac{\hbar}{2\omega_\alpha}}(\hat{a}^\dagger_\alpha + \hat{a}_\alpha)$ is the photonic displacement-field coordinate, related to the electric displacement operator, while $\hat{p}_\alpha$ is proportional to the magnetic field. The electron-photon coupling strength $\lambda_\alpha$ generally depends on the mode function of the cavity, but here we will take it as a constant, assuming that the cavity is much longer than the spatial range of the molecular dynamics.
The dipole self-energy term $\hat{V}_{\rm dipSE} = \frac{1}{2}\left(\lambda_\alpha(R - r)\right)^2$, has a negligible effect in all cases studied.  Polaritonic surfaces, defined by the eigenvalues of $\hat{H} - \hat{T}_n$, are shown in the lower panel of Fig.~\ref{fig:bopolpcet} for  coupling strengths $\lambda = 0.005$a.u. and 0.001a.u., and cavity-frequency $\omega_\alpha = 0.1$a.u. Immediately evident is the increased number of avoided crossings compared to the BO surfaces, as non-adiabatic effects from photon-matter and electron-nuclear couplings come into play. 

\begin{figure}
\includegraphics[width=0.9\columnwidth]{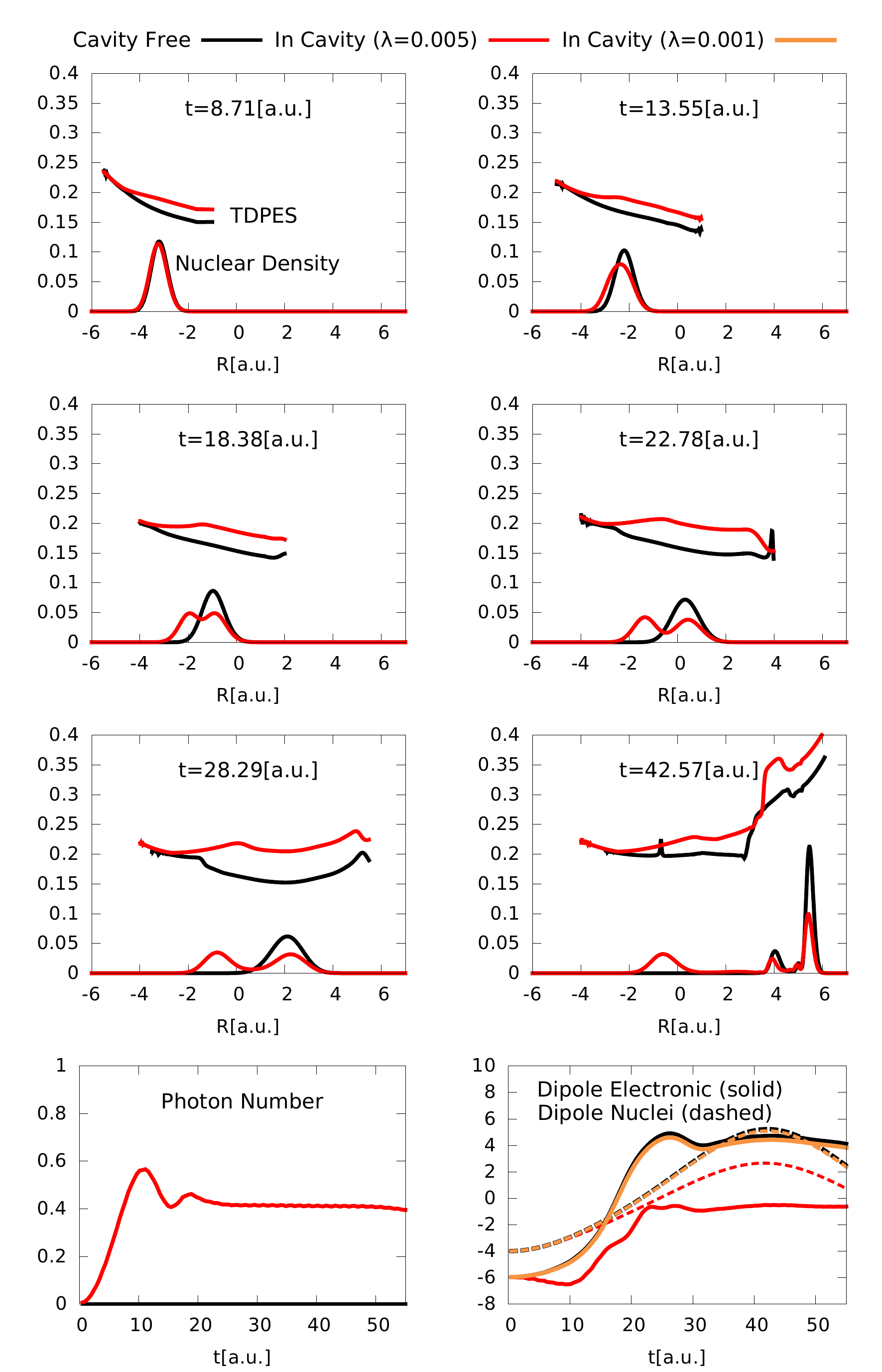}
\caption{Snapshots of the nuclear density (scaled by 0.1) and exact TDPES for dynamics inside (red) and outside (black) the cavity. The lowest panels show the electronic and nuclear dipole moments and the photon number over time. }
\label{fig:tdpes-nd-dips}
\end{figure}

Turning to the dynamics, Figure~\ref{fig:tdpes-nd-dips} shows time-snapshots of the nuclear density (red) resulting from the initial wavefunction, $\Psi(r,q,R,t) = \mathcal{N}e^{-\frac{(R+4)^2}{2.85}}\Phi_R^{\rm BO,(2)}(r)\xi^{(0)}(q)$, where $\xi^{(0)}(q) = (\omega_\alpha/\pi)^\frac{1}{4}e^{-\omega_\alpha q^2/2}$ is the zero-photon state in the cavity. The figure and the movie in Supplemental Material, demonstrate cavity-induced suppression of PCET: Significantly less proton density moves to the right compared to the cavity-free case (black), and while the electron transfer is clearly in concert with the proton transfer in the cavity-free dynamics as indicated by the black dipoles shown in the lower right panel, it is partially suppressed when the molecule is placed in the cavity with coupling strength $\lambda=0.005$a.u. The snapshots show that part of the molecular wavepacket becomes trapped in the donor well, reducing the nuclear dipole moment, consequently reducing the electron-transfer.  

Attempting to understand the suppression from the shape of the polaritonic surfaces of Fig.~{\ref{fig:bopolpcet}} alone is impossible: one might be tempted to attribute the partial trapping of the density to the barrier in the 3rd polaritonic surface at around $R \approx -2$a.u., however not only does the trapped portion of the density evolve past this point, but also the barrier is present in the weaker coupling $\lambda =0.001$a.u. case which shows negligible suppression as indicated by the orange dipole shown in the lower panel; the Supplemental Material also provides a movie of the density for this case. Instead, as we will shortly discuss, the structure of the exact TDPES shown in Fig.~\ref{fig:tdpes-nd-dips} directly correlates with the dynamics.

The TDPES is a fundamental construct arising from the EF approach~\cite{AMG10,AMG12}. When extended to systems of coupled electrons, nuclei, and photons~\cite{HARM18,AKT18}, this approach is based on factorizing the exact coupled wavefunction  into a nuclear wavefunction $\chi(R,t)$ and a conditional electron-photon wavefunction $\Phi_R(r,q,t)$, 
$\Psi(r,q,R,t) = \chi(R,t)\Phi_R(r,q,t)$, in which the exact equation for the marginal $\chi(R,t)$ has a Schr\"odinger form,
\ben
\left(-(\nabla + A(R,t))^2/2M+ \epsilon(R,t)\right)\chi(R,t) = i \partial_t\chi(R,t)
\een
(written here for one nuclear coordinate), 
 with a scalar potential $\epsilon(R,t)$, referred to as the TDPES, and a vector potential $A(R,t)$, both of which depend on the conditional electron-photon wavefunction. The time-evolution for the latter has a far more complicated form~\cite{GLM19}, with a non-Hermitian operator that operates on the $R$-dependence of $\Phi_R(r,q,t)$ and depends on the nuclear wavefunction $\chi(R,t)$. The exact equations are fully provided in the Supplemental Material.  The roles of the nuclei, electrons, and photons can be permuted in the factorization such that the subsystem of most interest is chosen for the marginal factor $\chi$ since this satisfies the Schr\"odinger equation~\cite{HARM18}, e.g. choosing the photonic system as the marginal, Ref.~\cite{HARM18} found distortions of the exact potential driving the photonic field away from harmonic due to photon-matter coupling.

 The factorization of $\Psi$ is unique up to an $(R,t)$-dependent phase-factor multiplying $\chi(R,t)$ with its inverse multiplying $\Phi_R(r,q,t)$; this in turn transforms the potentials, and in the case of one nuclear dimension, a gauge can always be found in which $A(R,t)$ is zero. 
In this gauge, the only potential driving the nuclei is $\epsilon(R,t)$ and, for the cavity-enclosed PCET model, this is shown in the time-snapshots of Fig.~\ref{fig:tdpes-nd-dips}. Comparing with the cavity-free TDPES, the structures that lead to the partial trapping of the nuclear density, and the subsequent partial suppression of PCET, are clearly seen. At early times, the slope of the TDPES is smaller compared to the cavity-free case, even sloping upwards in the trailing part of wavepacket, therefore slowing down  and spreading out the wavepacket compared to the cavity-free  case (up to $t=13.55$a.u.). A gentle step develops, lowering the potential on the left of the wavepacket, which begins to split the wavepacket in two parts ($t=18.38$a.u.); one part  becomes associated with TDPES turning downwards and forming a well to the left and the other turning downwards to the right, further enhancing the splitting. The nuclear wavepacket on the left is trapped in the well, and eventually will oscillate in it.  In contrast, the nuclear wavepacket on the right continues moving to the right ($t=22.78,28.29$a.u.), where it later splits, and behaves similarly to the cavity-free dynamics however scaled down due to having lost some density to the trapped region on the left ($t=42.57$a.u.). 

\begin{figure}
\includegraphics[width=0.9\columnwidth]{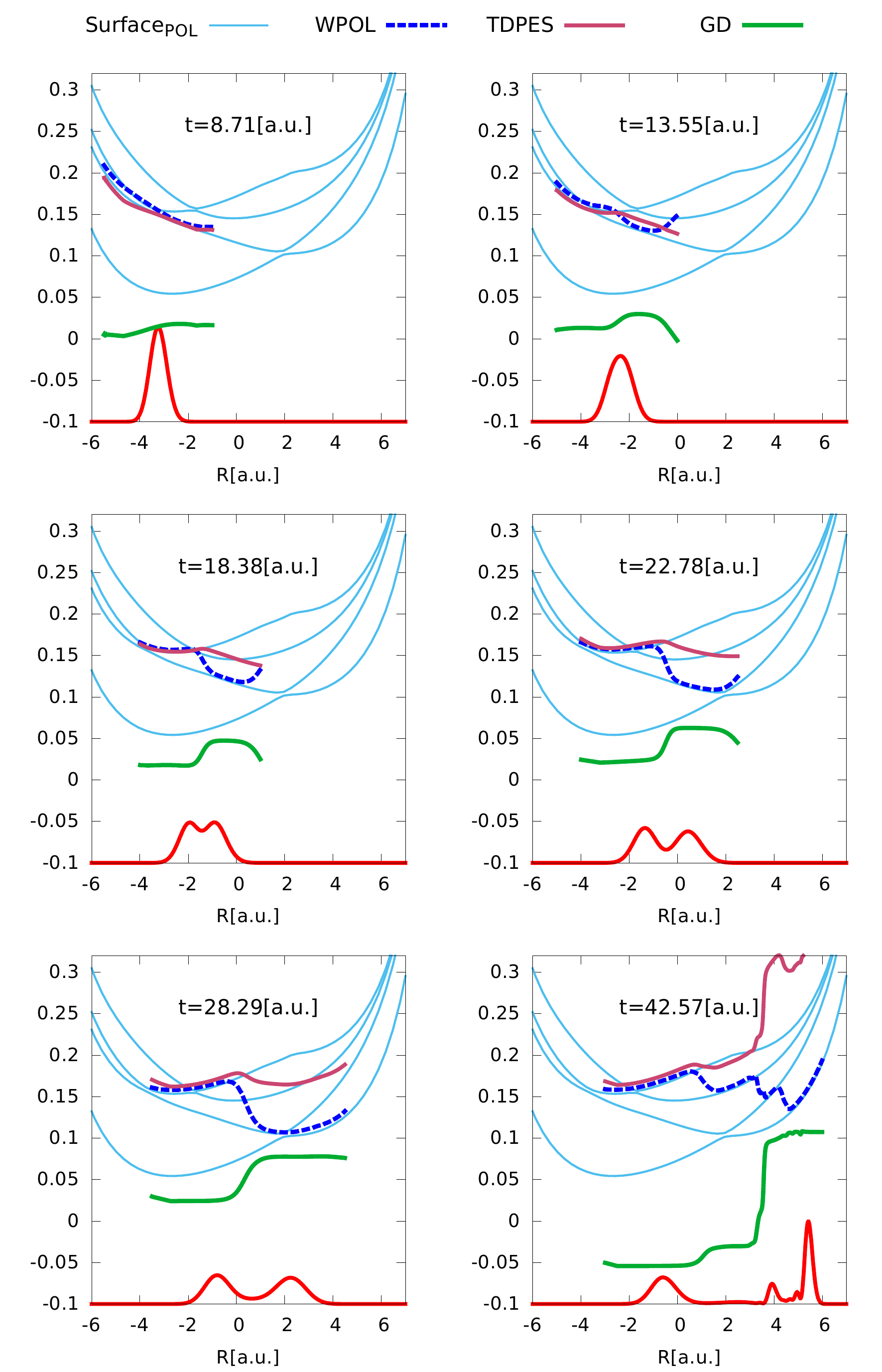}
\caption{Snapshots of the nuclear density and the components of the TDPES for dynamics in the cavity, $\lambda = 0.005, \omega_\alpha = 0.1$.  The thin background lines represent the polaritonic surfaces.}
\label{fig:tdpes-cpts}
\end{figure}

\begin{figure}
\includegraphics[width=0.8\columnwidth,height=\columnwidth]{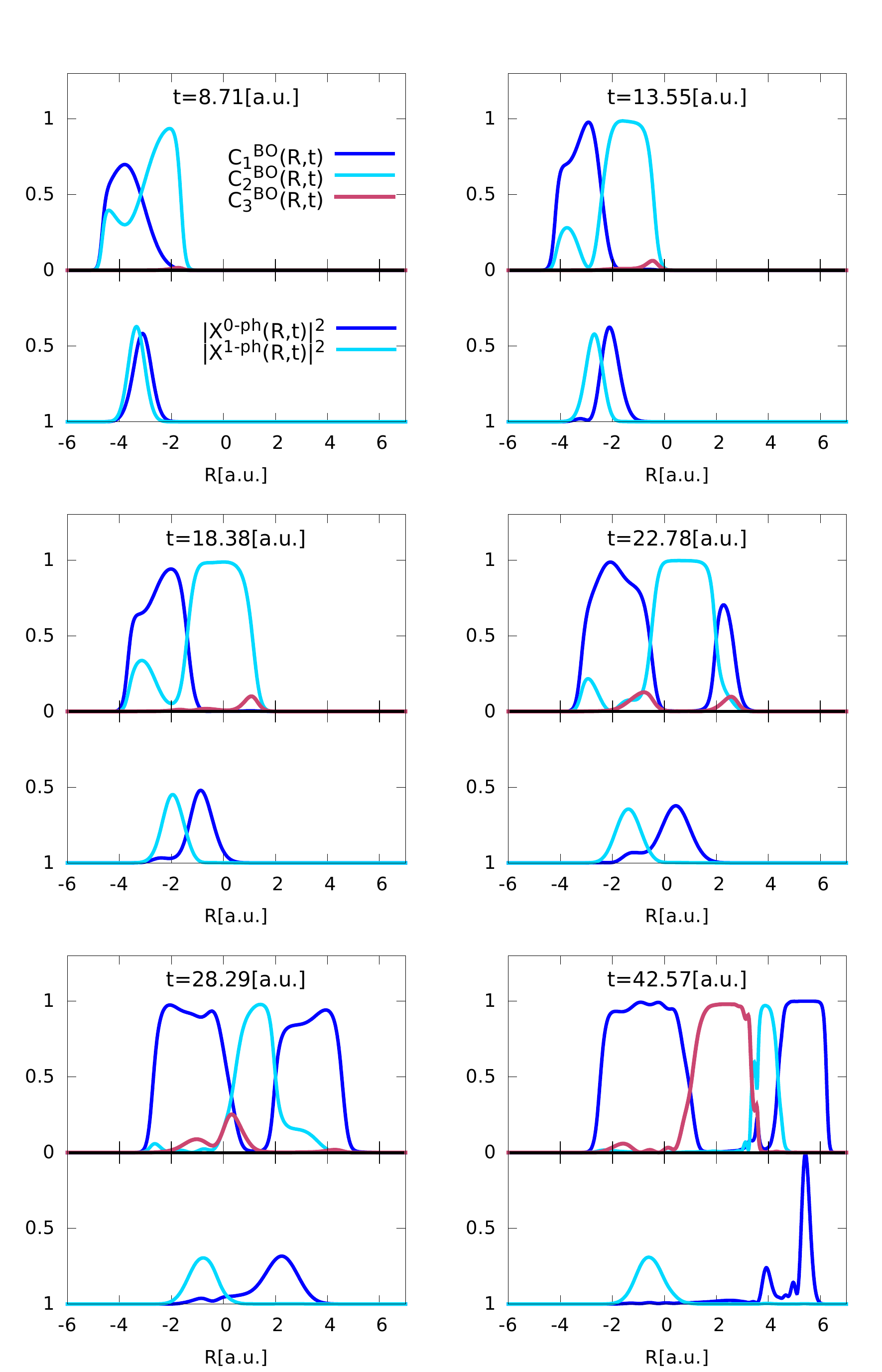}
\caption{Snapshots of the $0$-and $1$-photon resolved nuclear densities of Eq.~\ref{eq:nphotondens} (lower panel), along with the BO-coefficients of Eq.~\ref{eq:CBO} (upper panel)}
\label{fig:nphotondens_CBO}
\end{figure}

The shape of the TDPES therefore directly reflects the dynamics of the proton, but to understand the physical mechanisms yielding its shape, we consider the TDPES against the backdrop of polaritonic surfaces. First, we decompose the surface into weighted polaritonic (wpol), kinetic (kin), and gauge-dependent (GD) components that naturally arise from the form of the EF~\cite{AMG12,HARM18} (see also Supplemental Material):
\bea
\eps(R,t) &=& \eps_{\rm wpol}(R,t) + \eps_{\rm kin}(R,t) + \eps_{\rm GD}(R,t)\\
\eps_{\rm wpol}(R,t) &=& \langle \Phi_R \vert \hat{H}_{\rm BO} + \hat{H}_p + \hat{V}_{pm} \vert\Phi_R\rangle_{r,q}\\
\eps_{\rm kin}(R,t) &=& \langle \Phi_R \vert -\nabla^2_R\Phi_R\rangle_{r,q}/2M\\
\eps_{\rm GD}(R,t) &=& \langle \Phi_R\vert -i \partial_t\Phi_R\rangle_{r,q}
\eea
In Fig.~\ref{fig:tdpes-cpts} we plot $\eps_{\rm wpol}(R,t)$ and $\eps_{\rm GD}(R,t)$ against the backdrop of the static polaritonic surfaces; $\eps_{\rm kin}$ remains negligible throughout the evolution, due to the large proton mass.  At early times we observe that $\eps_{\rm wpol}$  on the left lies intermediate between the second and third polaritonic surfaces, acquiring a mixed character, while on the right adheres to the second polaritonic surface.  Looking at the middle row, this behavior resolves clearly into the left part of the nuclear wavepacket being correlated with the third polaritonic surface, while the right correlates with the second: this piecewise behavior illustrates the matter-photon correlation, with the left part correlated with photon-emission accompanying an electronic transition to the lower BO surface (see also Fig.~\ref{fig:nphotondens_CBO} shortly), while the right part of the nuclear wavepacket is correlated with a zero-photon electronically-excited state as in the initial state.
The step in $\eps_{\rm wpol}$ that bridges the two polaritonic surfaces after the photon-emission event is analogous to that found in earlier work between BO surfaces~\cite{AASG13} and between Floquet surfaces~\cite{FHGS17}, which polaritonic surfaces reduce to in the classical-light limit~\cite{GMDJ97}.
Also, analogous is that $\eps_{\rm GD}$ displays a countering step~\cite{AASMMG15}, that  provides a "realignment" that adjusts the energy locally in the nuclear system to account for the different energies of the electron-photon system associated with the different characters on the left and right. But it is important to note that the suppression mechanism sets in earlier, during the stage when the surface has a mixed character, before the shifted-piecewise character of $\eps_{\rm GD}$ sets in. This is also well-before part of the wavepacket encounters  the  avoided crossing associated with strong electron-nuclear coupling around $R\approx 2$a.u. (see also the BO surfaces in Fig. 1), which is where the nuclear wavepacket splits again with the part that moves to the lowest surface associated with the electron-transfer. This latter splitting also occurs for the cavity-free case as we saw in Fig. 2.  At the final time shown we see three parts to the nuclear wavepacket: the left part trapped in the well on the right associated with a 1-photon BO ground-state, and two lobes on the right, with the extreme right associated with the PCET on the BO ground-state, and the other associated with the electronically excited BO state, both with 0 photons. The component of the exact TDPES $\eps_{\rm wpol}$ directly reflects this matter-polariton correlation, while $\eps_{\rm GD}$ adjusts the local energy in a piecewise manner. 

To further clarify the dynamics in the conditional variables $q$ and $r$, Figure~\ref{fig:nphotondens_CBO} shows the $n$-photon resolved nuclear density, defined as
\ben
\vert\chi^{n-ph}(R,t)\vert^2 =\vert \langle\xi_n\vert\Psi(t)\rangle_{r,q} \vert^2
\label{eq:nphotondens}
\een
where $\xi_n(q)$ are the harmonic oscillator eigenstates of $H_{p}$, and the BO-coefficients are defined as 
\ben
C_{i}^{\rm BO}(R,t) = \vert\langle \Phi_R^{{\rm BO}, i} \vert \Psi(t)\rangle_{r,q}\vert^2
\label{eq:CBO}
\een
These measures  very clearly show the nuclear-photon and nuclear-electron correlations throughout the evolution (see also a movie in the Supplemental Material). At early times we see the mixed character of the electron-photon state, with both 0-photon and 1-photon contributions associated with the nuclear density at a given $R$, and fractional BO coefficients contributing (with even the third BO state being appreciably occupied). Only after the photon-emission event is the nuclear density locally correlated with only one electronic or photonic state. 

In conclusion, we have analyzed the structure of the exact TDPES for a model of PCET, , and shown how its features can predict the suppression induced by the cavity. While the polaritonic surfaces themselves provide a very useful backdrop, they are themselves not able to predict the dynamics or mechanisms without considering how they couple to each other in a dynamics scheme~\cite{KBM16,KM17}, and care is needed with such a scheme, due to the propensity of near-crossings caused by both electron-nuclear and electron-photon coupling. As a result, for mixed quantum-classical methods which would be required for many-molecule systems~\cite{LFTG17,MC19} over-coherence in surface-hopping methods is likely to be more problematic. 
Instead, this work shows the promise of rigorously-based mixed quantum-classical approximations for cavity-qed, based on generalizations of the coupled-trajectory scheme of Ref.~\cite{AMAG16,MAG15,MATG17,AC19,GAM18,HLM18}, for example, that has been successful for cavity-free non-adiabatic dynamics. 


\begin{acknowledgments}{Financial support from the US National Science Foundation
CHE-1566197 (L.L.) and the Department of Energy, Office
of Basic Energy Sciences, Division of Chemical Sciences,
Geosciences and Biosciences under Award DE-SC0015344 (N.T.M)
are gratefully acknowledged. NMH gratefully acknowledges funding from the Max-Planck Institute for the Structure and Dynamics of Matter, and an IMPRS fellowship.}
\end{acknowledgments}
\bibliography{./ref_na}

\end{document}


\title{Supplementary Material for "Exact Potential Energy Surface for Molecules in Cavities"}
\author{Lionel Lacombe}
\affiliation{Department of Physics and Astronomy, Hunter College of the City University of New York, 695 Park Avenue, New York, New York 10065, USA}
\author{Norah M. Hoffman}
\affiliation{Max Planck Institute for the Structure and Dynamics of Matter and
Center for Free-Electron Laser Science and Department of Physics,
Luruper Chaussee 149, 22761 Hamburg, Germany}
\affiliation{Department of Physics and Astronomy, Hunter College of the City University of New York, 695 Park Avenue, New York, New York 10065, USA}
\author{Neepa T. Maitra}
\affiliation{Department of Physics and Astronomy, Hunter College  of the City University of New York, 695 Park Avenue, New York, New York 10065, USA}
\affiliation{Physics Program and Chemistry Program, Graduate Center of the City University of New York, New York, USA}

\date{\today}
\pacs{}
\begin{abstract}
\end{abstract}
\maketitle

\section{Exact Factorization Equations}
The full equations of the EF factorization we use here are based on the generalization given in Ref.~\cite{HARM18} of the original time-dependent EF equations of Ref.~\cite{AMG10,AMG12}. 
These are based on the factorization
\ben
\Psi(\dulr,\dulq,\dulR,t) = \chi(\dulR,t)\Phi_\dulR(\dulr,\dulq,t)
\een
where $\dulr,\dulq,\dulR$ represent all electronic-, photonic displacement-field mode-, and nuclear- coordinates, respectively, and the partial normalization condition
\ben
\int\vert\Phi_\dulR(\dulr,\dulq,t)\vert^2 d\dulr d\dulq = 1
\een
is satisfied.
The marginal and conditional parts each satisfy the following coupled equations of motion:
\bea
\label{eq:Phi}
\left(\hat{H}_ {\rm BO} + \hat{H}_p +\hat{V}_{pm} + \hat{V}_{\rm dipSE} + \hat{U}_{\rm ep-n} - \eps(\dulR,t)\right)\Phi_\dulR(\dulr,\dulq,t) &=& i \partial_t  \Phi_\dulR(\dulr,\dulq,t)\\
\left(\sum_{J=1}^{N_n}( -i\nabla_J + \bA_J(\dulR,t))^2/2M_J + \eps(\dulR,t)\right)\chi(\dulR,t) &=& i\partial_t\chi(\dulR,t)
\label{eq:chi}
\eea
with 
\bea
\hat{U}_{ep-n} &=&\sum_{J=1}^{N_n}\frac{1}{M_J}\left( \frac{(-i\nabla_J - \bA_J(\dulR,t))^2}{2} + \left(\frac{-i\nabla_J\chi(\dulR,t)}{\chi(\dulR,t)} +\bA_J(\dulR,t)\right)\cdot\left(-i\nabla_J - \bA_J(\dulR,t)\right)\right)\\
\eps(\dulR,t) &=& \langle\Phi_\dulR\vert\hat{H}_ {\rm BO} + \hat{H}_p +\hat{V}_{pm} + \hat{V}_{\rm dipSE} + \hat{U}_{\rm ep-n} -i\partial_t \vert\Phi_\dulR\rangle_{\dulr,\dulq}\\
\bA_J(\dulR,t) &=& \langle \Phi_\dulR\vert -i\nabla_J \Phi_\dulR\rangle_{\dulr,\dulq}
\label{eq:A}
\eea
and all other terms in Eqs.~\ref{eq:Phi} and~\ref{eq:chi} are given in the main text for the one-dimensional model we studied. The notation $\langle ...\rangle_{\dulr,\dulq}$ indicates an integral over all photonic displacement-field and electronic coordinates only.

The marginal part, $\chi(\dulR,t)$ is a nuclear wavefunction in the sense that it reproduces the exact nuclear density and exact nuclear current-density of the exact full photon-matter wavefunction. The equations \ref{eq:Phi}--\ref{eq:A} are form-invariant under the phase-transformation $\Phi_{\dulr,\dulq}(\dulR,t) \to \Phi_{\dulr,\dulq}(\dulR,t)e^{i\theta(\dulR,t)}, \chi(\dulR,t) \to \chi(\dulR,t) e^{-i\theta(\dulR,t)}$ with the potentials undergoing the gauge-like transformation $\bA_J(\dulR,t) \to \bA_J(\dulR,t) + \nabla_J\theta(\dulR,t), \eps(\dulR,t) \to \eps(\dulR,t) + \partial_t\theta(\dulR,t)$, and the factorization Eq. 1 is unique up to such a transformation.

The model we studied has a one-dimensional nuclear coordinate so a gauge can always be found in which the vector potential $\bA(\dulR,t)$ is zero. This is the gauge we chose in our calculations. The equations then simplify in the sense, for example, that there is only one potential, the scalar $\eps(\dulR,t)$ appearing in the nuclear equation, and the scalar potential can then be written as three terms, as prescribed in Eqs. (6)--(9) of the main paper. 

In practise, we obtained the potential energy surface $\eps(\dulR,t)$ by inversion~\cite{AMG12}. That is, we first solved the time-dependent Schr\"odinger equation for $\Psi(r,q,R,t)$ on a three-dimensional grid, and extracted $\chi(R,t) = \vert\chi(R,t)\vert e^{iS(R,t)}$ using 
\ben
\vert\chi(R,t)\vert = \sqrt{\int dq dr \vert\Psi(r,q,R,t)\vert^2}
\een
 and 
 \ben
 S(R,t) = \int^R\left( \frac{{\rm Im} \int dr dq\,\Psi(r,q,R',t)\frac{d\Psi(r,q,R',t)}{dR'}}{\vert\chi(R',t)\vert^2} \right) dR'
\een
Then we found $\Phi_{r,q}(R,t) = \Psi(r,q,R,t)/\chi(R,t)$ enabling us to evaluate the matrix elements involved for $\eps(R,t)$ (Eqs.(6) -- (9) of the main text). 

\section{Numerical details}
In our calculations, we used 192, 96, 1280 points on a grid of size $\pm 120.20$ a.u., $\pm 80$ a.u., $\pm 9.5$ a.u., for the electronic, photonic or nuclear calculation respectively. 
We also used a time-step of $0.1$ a.u.

\section{Movies}
We provide three movies:

(i) {\bf movieCpl0p005} shows the dynamics for the case where the resonant frequency of the cavity is $\omega_\alpha  = 0.1$au and the coupling strength is $\lambda = 0.005$ in red, compared to the cavity-free dynamics in black.  

Top left: exact TDPES, shown against the background of polaritonic surfaces in blue and green, and the nuclear density, scaled by 0.1 and shifted down, is shown in the lower part of all plots in the first row. 

Top middle: weighted polaritonic component of the exact TDPES $\eps_{\rm wpol}(R,t)$ .

Right middle: gauge-dependent component of the exact TDPES $\eps_{\rm GD}(R,t)$. 

Lower left: BO coefficients $C_i(R,t)$ as defined in Eq. (11) of the main text as a function of time

Lower middle: number of photons emitted as a function of time

Lower right: electronic (solid) and nuclear (dashed) dipole moments as a function of time.

\vspace{0.4cm}
(ii) {\bf movieCpl0p001} as above but for coupling strength $\lambda = 0.001$. 

\vspace{0.4cm}
(iii) {\bf movieCpl0p005phdenCBO} shows the $n$-photon resolved densities and BO-coefficients for the $\omega_\alpha =0.1$ and $\lambda = 0.005$ case as compared with the cavity-free case. 

Top left panel: the total nuclear density (as a reference)

Top right panel: the $0$-photon resolved density

Middle left: the $1$-photon resolved density

Middle right: the $2$-photon resolved density

Lower left: the BO coefficients in the cavity

Lower right: the BO-coefficients for the cavity-free case.
\bibliography{./ref_na}